\documentclass{svmult}

\usepackage{graphicx}    

\newcommand{\dddot}[1]{\stackrel{\ldots}{#1}}
\def\mathbfepsilon{\mathbf{\epsilon}}
\newcommand{\refb}[1]{(\ref{#1})}
\newcommand{\thickspace}{\;}

\def\beq{\begin{equation}}
\def\eeq{\end{equation}}

\def\bea{\begin{eqnarray}}
\def\eea{\end{eqnarray}}
\def\ba{\begin{array}}                  
\def\ea{\end{array}}

\begin{document}

\title*{Microscopic interpretation of black hole entropy}

\author{
M. Cvitan\inst{1}\and
S. Pallua\inst{2}\and
P. Prester\inst{3}}
\institute{
\texttt{E-mail: mcvitan@phy.hr}
\and 
\texttt{E-mail: pallua@phy.hr}
\and 
\texttt{E-mail: pprester@phy.hr}\\
Theoretical Physics Department, \\
Faculty of Natural Sciences and Mathematics, \\
University of Zagreb, \\
Bijeni\v{c}ka c. 32, pp. 331, HR--10002 Zagreb, 
  Croatia\\
}
%
%
\maketitle

\abstract{It is shown, using conformal symmetry methods, that one can 
obtain  microscopic interpretation of black hole entropy for general 
class of higher curvature Lagrangians.}

\section{Introduction}

The entropy of black holes can be calculated with the well known 
Bekenstein-Hawking formula
\begin{equation}\label{e1}
S_{BH} = \frac{A}{4\pi G}
\thickspace\textrm{,}
\end{equation}
where $A$ represents area of black hole horizon. In fact 
generalisation of this formula is given in \cite{Iyer:1994ys}
for general interaction of the form
\begin{equation}\label{e2}
L = L(g_{ab}, R_{abcd}, \nabla R_{abcd}, \psi, \nabla \psi,...)
\thickspace\textrm{.}
\end{equation}
Here $\psi$ refers to matter fields and dots refer to derivatives up
to order $m$. In that case the entropy is given with the relation
\cite{Iyer:1994ys}
\begin{equation}\label{e3}
S=-2 \pi \int_{\mathcal{H}\cap C} \hat{\epsilon} E^{abcd} \eta_{ab} \eta_{cd}
\thickspace\textrm{.}
\end{equation}
Here $\mathcal{H}\cap C$ is a cross section of the horizon, $\eta_{ab}$ denotes 
binormal to $\mathcal{H}\cap C$, $\hat{\epsilon}$ is induced volume
element on $\mathcal{H}\cap C$ 
and
\begin{equation}\label{e4}
E^{abcd} = \frac{\partial L}{\partial R_{abcd}} - \nabla_{a_{1}} 
\frac{\partial L}{\partial \nabla_{a_{1}}R_{abcd}} + \ldots (-)^m 
\nabla_{(a_{1} \ldots a_{m})} \frac{\partial L}{\partial 
\nabla_{(a_{1} \ldots a_{m})} R_{abcd}}
\thickspace\textrm{.}
\end{equation}
The problem of microscopic description of black hole entropy was 
approached by different methods like string theory which treated 
extremal black holes \cite{Strominger:1996sh} or e.g.~loop 
quantum gravity \cite{Ashtekar:1997yu}. An interesting line of 
approach is based on conformal field theory and Virasoro algebra. One
particular formulation was due to Solodukhin who reduced the problem
of $D$-dimesional black holes to effective two-dimensional theory with
fixed boundary conditions on the horizon. The effective theory was 
found to admit Virasoro algebra near horizon. Calculation of its 
central charge allows then to compute the entropy 
\cite{Solodukhin:1998tc,Cvitan:2002cs}. An independent formulation is
due to Carlip 
\cite{Carlip:2002be,Carlip:2001kk,Carlip:1998wz,Carlip:1999cy}
who has shown that under certain simple assumptions on boundary 
conditions near black hole horizon one can identify a subalgebra of 
algebra of diffeomorphisms which turns out to be Virasoro algebra. The
fixed boundary conditions give rise to central extensions of this 
algebra. The entropy is then calculated from Cardy formula 
\cite{Cardy:ie}
\begin{equation}\label{e5}
S_{c}= 2\pi\sqrt{(\frac{c}{6}-4\Delta_{g})(\Delta-\frac{c}{24})}
\thickspace\textrm{.}
\end{equation}
Here $\Delta$ is the eigenvalue of Virasoro generator $L_0$ for the 
state we calculate the entropy and $\Delta_{g}$ is the smallest 
eigenvalue. In that way the entropy formula \refb{e1} for Einstein 
gravity was reproduced. In this lecture we want to investigate if such 
microscopic interpretation is possible for more general type of 
interaction. We shall first treat Gauss-Bonnet gravity using Solodukhin method and then using Carlip method. The latter method will allow us to treat  more general cases. These are described   with Lagrangian  which is allowed to have arbitrary dependence on Riemann tensor but not on its derivatives, more precisely
\begin{equation}\label{e6}
L = L(g_{ab}, R_{abcd})
\thickspace\textrm{.}
\end{equation}
In that case the tensor $E^{abcd}$ takes the form
\begin{equation}\label{e7}
E^{abcd} = \frac{\partial L}{\partial R_{abcd}}
\thickspace\textrm{.}
\end{equation}
We note that interesting new posibilities and open questions arise for interpretation of black hole entropy.
For discussion in the Gauss--Bonnet case see e.g.\ 
\cite{Nojiri:2001ae,Cvetic:2001bk,Nojiri:2002hz,Cho:2002hq,Neupane:2002bf,Clunan:2004tb}.

\section{Effective CFT near the horizon}

Now we turn our attention to particular microscopic derivation of
entropy of black hole, which was first done in 
\cite{Solodukhin:1998tc} for the Einstein gravity, and then extended
to general D-dimensional Gauss--Bonnet (GB) theories in
\cite{Cvitan:2002cs}. General GB action\footnote{Also known as 
Lovelock gravity.} is given by
\begin{equation}\label{igb}
I_{\mbox{\scriptsize GB}}=
-\sum_{m=0}^{[D/2]}\lambda_m\int d^D x\sqrt{-g}\mathcal{L}_m(g)\;,
\end{equation}
where GB densities $\mathcal{L}_m(g)$ are
\begin{equation}\label{lgbm}
\mathcal{L}_m(g)=\frac{(-1)^m}{2^{m}}\delta_{\mu_1\nu_1\ldots
\mu_m\nu_m}^{\rho_1\sigma_1\ldots\rho_m\sigma_m}
{R^{\mu_1\nu_1}}_{\rho_1\sigma_1}\cdots
{R^{\mu_m\nu_m}}_{\rho_m\sigma_m}\;,
\end{equation}
We take $\lambda_0=0$ (cosmological constant), because we shall see 
that this term is irrelevant for our calculation. Coupling constant 
$\lambda_1$ is related to more familiar $D$-dimensional Newton 
gravitational constant $G_D$ through $\lambda_1=(16\pi G_D)^{-1}$.

We neglect matter and consider
$S$-wave sector of the theory, i.e., we consider only radial
fluctuations of the metric. It is easy to show that in this case
(\ref{igb}) can be written in the form of an effective
two-dimensional ``generalised higher-order Liouville theory'' given 
with
\begin{eqnarray} \label{gbef}
I_{\mbox{\scriptsize GB}}&=&\Omega_{D-2}\sum_{m=0}^{[D/2]}\lambda_m
\frac{(D-2)!}{(D-2m)!}\int d^2x\sqrt{-\gamma}\, r^{D-2m-2}
\left[1-(\nabla r)^2\right]^{m-2} \nonumber \\ &&\times \bigg\{
2m(m-1)r^2\left[(\nabla_a\nabla_br)^2-(\nabla^2r)^2\right] \nonumber\\
&&\quad+2m(D-2m)r\nabla^2r\left[1-(\nabla r)^2\right]
+m\mathcal{R}r^2\left[1-(\nabla r)^2\right] \nonumber \\
&&\quad\left. -(D-2m)(D-2m-1)\left[1-(\nabla r)^2\right]^2\right\}\;.
\end{eqnarray}

We now suppose that black hole with horizon \emph{is existing} and we
are interested in fluctuations (or, better, quantum states) near it. In
the spherical geometry apparent horizon $\mathcal{H}$ (a line in
$x$-plane) can be defined by the condition \cite{Russo95}
\begin{equation}\label{hoco}
\left.(\nabla r)^2\right|_\mathcal{H}\equiv
\left.\gamma^{ab}\partial_a r\partial_b r\right|_\mathcal{H}=0 \;.
\end{equation}
Notice that (\ref{hoco}) is invariant under (regular) conformal 
rescalings of the effective two-dimensional metric $\gamma_{ab}$. Near
the horizon (\ref{hoco}) is approximately satisfied. It is easy to see 
that after partial integration and implementation of horizon condition
$(\nabla r)^2\approx 0$, (\ref{gbef}) becomes near the horizon
approximately
\begin{eqnarray} \label{gbefh}
I_{\mbox{\scriptsize GB}}&=&-\Omega_{D-2}\sum_{m=0}^{[D/2]}\lambda_m
\frac{(D-2)!}{(D-2m-2)!}\int d^2x\sqrt{-\gamma}\, r^{D-2m-2}\nonumber\\
&&\times\left\{m(\nabla r)^2-\frac{m}{(D-2m)(D-2m-1)}\mathcal{R}r^2
+1\right\} \;.
\end{eqnarray}
If we define
\begin{equation} \label{rFi}
\Phi^2\equiv 2\Omega_{D-2}\sum_{m=1}^{[D/2]}m\lambda_m
\frac{(D-2)!}{(D-2m)!}\, r^{D-2m} \;,
\end{equation}
and make reparametrizations
\begin{equation}\label{rfi}
\phi\equiv\frac{2\Phi^2}{q\Phi_h} \;, \qquad
\tilde{\gamma}_{ab}\equiv\frac{d\phi}{dr}\gamma_{ab}\;,
\end{equation}
where $q$ is arbitrary dimensionless parameter, the action
(\ref{gbefh})
becomes
\begin{equation} \label{gbef2}
I_{\mbox{\scriptsize GB}}=\int d^2x\sqrt{-\tilde{\gamma}}
\left[\frac{1}{4}q\Phi_h\phi\tilde{\mathcal{R}}-V(\phi)\right]
\end{equation}

This action can be put in more familiar form if we make additional
conformal reparametrization:
\begin{equation}
\bar{\gamma}_{ab}\equiv e^{-\frac{2\phi}{q\Phi_h}}
\tilde{\gamma}_{ab}\;,
\end{equation}
Now (\ref{gbef2}) takes the form
\begin{equation}\label{gb2ef3}
I_{\mbox{\scriptsize GB}}=-\int d^2x\sqrt{-\bar{\gamma}}
\left[\frac{1}{2}(\bar{\nabla}\phi)^2-
\frac{1}{4}q\Phi_h\phi\bar{\mathcal{R}}+U(\phi)\right]\;,
\end{equation}
which is simmilar to the Liouville action. The difference is that
potential $U(\phi)$ is not purely exponential which means that the
obtained effective theory is not exactly conformaly invariant. 

Action (\ref{gb2ef3}) is of the same form as that obtained from pure
Einstein action. In \cite{Solodukhin:1998tc} it was shown that if one
imposes condition that the metric $\bar{\gamma}_{ab}$ is 
\emph{nondynamical} then the action (\ref{gb2ef3}) describes CFT 
\emph{near the horizon}\footnote{Carlip showed that above condition is
indeed consistent boundary condition \cite{Carlip:2001kk}.}. We 
therefore fix $\bar{\gamma}_{ab}$ near the horizon and take it to be 
a metric of a static spherically symmetric black hole:
\begin{equation}\label{ds2w}
d\bar{s}_{(2)}^2\equiv\bar{\gamma}_{ab}dx^adx^b=
-f(w)dt^2+\frac{dw^2}{f(w)}\;,
\end{equation}
where near the horizon $f(w_h)=0$ we have
\begin{equation}\label{fwh}
f(w)=\frac{2}{\beta}(w-w_h)+O\left((w-w_h)^2\right)\;.
\end{equation}
We now make coordinate reparametrization $w\to z$
\begin{equation}\label{zw}
z=\int^w\frac{dw}{f(w)}=\frac{\beta}{2}\ln\frac{w-w_h}{f_0}+O(w-w_h)
\end{equation}
in which 2-dim metric has a simple form
\begin{equation}\label{ds2z}
d\bar{s}_{(2)}^2=f(z)\left(-dt^2+dz^2\right)\;,
\end{equation}
and the function $f$ behaves near the horizon ($z_h=-\infty$) as
\begin{equation}\label{fzh}
f(z)\approx f_0 e^{2z/\beta}\;,
\end{equation}
i.e., it \emph{exponentially} vanishes. It is easy to show that
equation of motion for $\phi$ which follows from Eqs. (\ref{gb2ef3}),
(\ref{ds2z}), (\ref{fzh}) is
\begin{equation}\label{eom}
\left(-\partial_t^2+\partial_z^2\right)\phi=\frac{1}{4}q\Phi_h
\bar{\mathcal{R}}f+fU'(\phi)\approx O\left(e^{2z/\beta}\right)\;,
\end{equation}
and that the ``flat'' trace of the energy-momentum tensor is
\begin{equation}\label{temt}
-T_{00}+T_{zz}=\frac{1}{4}q\Phi_h\left(-\partial_t^2+
\partial_z^2\right)\phi-fU(\phi)\approx O\left(e^{2z/\beta}\right)\;,
\end{equation}
which is exponentially vanishing near the horizon\footnote{Higher
derivative terms in (\ref{gbef}) make contribution to (\ref{temt})
proportional to $f(\nabla\phi)^2\approx o(\exp(2z/\beta))$.}. From
(\ref{eom}) and (\ref{temt}) follows that the theory of the scalar
field $\phi$ exponentially approaches CFT near the horizon.

Now, one can construct corresponding Virasoro algebra using standard
procedure. Using light-cone coordinates $z_\pm=t\pm z$ right-moving
component of energy--momentum tensor near the horizon is approximately
\begin{equation} \label{tempp}
T_{++}=(\partial_+\phi)^2 - \frac{1}{2}q\Phi_h\partial_+^2\phi + 
\frac{q\Phi_h}{2\beta}\partial_+\phi \;.
\end{equation}
It is important to notice that horizon condition (\ref{hoco}) implies
that $r$ and $\phi$ are (approximately) functions only of one
light-cone coordinate (we take it to be $z_+$), which means that only
one set of modes (left \emph{or} right) is contributing.

Virasoro generators are coefficients in the Fourier expansion of
$T_{++}$:
\begin{equation} \label{tfour}
T_n=\frac{\ell}{2\pi}\int_{-\ell/2}^{\ell/2}dz\, e^{i2\pi nz/\ell}
T_{++} \;,
\end{equation}
where we compactified $z$-coordinate on a circle of circumference
$\ell$.
Using canonical commutation relations it is easy to show that Poisson
brackets of $T_n$'s are given with
\begin{equation} \label{fvir}
i\{T_n, T_m\}_{\mbox{\scriptsize PB}}=(n-m)T_{n+m}
+\frac{\pi}{4}q^2\Phi_h^2\left(n^3+n
\left(\frac{\ell}{2\pi\beta}\right)^2\right)\delta_{n+m,0} \;.
\end{equation}
To obtain the algebra in quantum theory (at least in semiclassical
approximation) one replaces Poisson brackets with commutators using
$[\,,]=i\hbar\{\,,\}_{\mbox{\scriptsize PB}}$, and divide generators
by $\hbar$. From (\ref{fvir}) it follows that ``shifted'' generators
\begin{equation} \label{lntn}
L_n=\frac{T_n}{\hbar}+\frac{c}{24}\left(\left(
\frac{\ell}{2\pi\beta}\right)^2+1\right)\delta_{n,0} \;,
\end{equation}
where
\begin{equation} \label{cch}
c=3\pi q^2\frac{\Phi_h^2}{\hbar} \;,
\end{equation}
satisfy Virasoro algebra
\begin{equation} \label{vir}
[L_n, L_m]=(n-m)L_{n+m}+\frac{c}{12}\left(n^3-n\right)\delta_{n+m,0}
\end{equation}
with central charge $c$ given in (\ref{cch}).

Outstanding (and unique, as far as is known) property of the Virasoro
algebra is that in its representations a logarithm of the number of
states (i.e., entropy) with the eigenvalue of $L_0$ equal to $\Delta$ is
asymptoticaly given with Cardy formula (\ref{e5}). 
If we assume that in our case $\Delta_g=0$ in semiclassical
approximation (more precisely, $\Delta_g\ll c/24$), one can see that
number of microstates (purely quantum quantity) is in leading
approximation completely determined by (semi)classical values of $c$
and $L_0$. Now it only remains to determine $\Delta$. In a classical
black hole solution we have
\begin{equation} \label{rsol}
r=w=w_h+(w-w_h)\approx r_h+f_0 e^{2z/\beta}\;,
\end{equation}
so from (\ref{rfi}) and (\ref{rFi}) follows that near the horizon
$\phi\approx\phi_h$.
Using this configuration in (\ref{tfour}) one obtains $T_0=0$, which
plugged in (\ref{lntn}) gives
\begin{equation} \label{delta}
\Delta=\frac{c}{24}
\left(\left(\frac{\ell}{2\pi\beta}\right)^2+1\right)\;.
\end{equation}
Finally, using (\ref{cch}) and (\ref{delta}) in Cardy formula
(\ref{e5}) one obtains
\begin{equation} \label{sconf}
S_{\mbox{\scriptsize C}}=\frac{c}{12}\frac{\ell}{\beta}
=\frac{\pi}{4}q^2\frac{\ell}{\beta}\frac{\Phi_h^2}{\hbar}\;.
\end{equation}
Let us now compare (\ref{sconf}) with classical formula (\ref{e3}),
which for GB gravities can be written as \cite{JacMye93}
\begin{equation}\label{sgb}
S_{\mbox{\scriptsize GB}}=\frac{4\pi}{\hbar}\sum_{m=1}^{[D/2]}
m\lambda_m\oint d^{D-2}x\sqrt{\tilde{g}}
\mathcal{L}_{m-1}(\tilde{g}_{ij})\;,
\end{equation}
Here $\tilde{g}_{ij}$ is induced metric on the horizon, and densities
$\mathcal{L}_{m}$ are given in (\ref{lgbm}). In the
sphericaly symmetric case horizon is a $(D-2)$-dimensional sphere with
radius $r_h$ and $R(\tilde{g}_{ij})=-(D-2)(D-3)/r_h^2$, so
(\ref{sgb}) becomes
\begin{equation} \label{sgbss}
S_{\mbox{\scriptsize GB}}=\frac{4\pi}{\hbar}\Omega_{D-2}
\sum_{m=1}^{[D/2]}m\lambda_m\frac{(D-2)!}{(D-2m)!}\, r^{D-2m}
=2\pi\frac{\Phi_h^2}{\hbar}
\end{equation}
Using this our expression (\ref{sconf}) can be written as
\begin{equation} \label{scsgb}
S_{\mbox{\scriptsize C}}=
\frac{q^2}{8}\frac{\ell}{\beta}S_{\mbox{\scriptsize GB}}\;,
\end{equation}
so it gives correct result apart from dimensionless coeficient, which
can be determined in the same way as in pure Einstein case
\cite{Carlip:2001kk}. First, it is natural to set the compactification
period $\ell$ equal to period of Euclidean-rotated black
hole\footnote{We note that our functions depend only on variable
$z_+$, so the periodicity properties in time $t$ are identical to
those in $z$.}, i.e.,
\begin{equation} \label{lwick}
\ell=2\pi\beta \;.
\end{equation}
The relation between eigenvalue $\Delta$ of $L_0$ and $c$ then becomes
\begin{equation} \label{delc}
\Delta=\frac{c}{12} \;.
\end{equation}
One could be tempted to expect this relation to be valid for larger 
class of black holes and interactions then those treated so far.

To determine dimensionless parameter $q$ we note that our effective
theory given with (\ref{gb2ef3}) depends on effective parameters
$\Phi_h$ and $\beta$, and thus one expects that $q$ depends on
coupling constants only through dimensionless combinations of them.
Thus to determine $q$ one may consider $\lambda_2=0$ case and compare
expression for central charge (\ref{cch}) with that obtained in
\cite{Carlip:1999cy}, which is
\begin{equation} \label{cchcar}
c=\frac{3A_h}{2\pi\hbar G_D}\;,
\end{equation}
where $A_h=\Omega_{D-2}r_h^{D-2}$ is the area of horizon. One obtains
that
\begin{equation} \label{qcar}
q^2=\frac{4}{\pi} \;.
\end{equation}
One could also perform boundary analysis of Ref. \cite{Carlip:1999cy} 
for GB gravity (see Appendix of \cite{Cvitan:2002cs}). This procedure
gives $\Delta=\Phi_h^2/\hbar$ which combined with (\ref{cch}) and
(\ref{delc}) gives (\ref{qcar}).

Using (\ref{lwick}) and (\ref{qcar}) one finally obtains desired
result
\begin{equation} 
S_{\mbox{\scriptsize C}}=S_{\mbox{\scriptsize GB}} \;.
\end{equation}

Let us mention that there were other approaches to calculation of
black hole entropy using Virasoro algebra of near-horizon symmetries
of effective 2-dim QFT (see \cite{GiaPin}).

\section{Covariant phase space formulation of gravity}

As mentioned before there is another method in which one is not using
dimensional reduction. The emphasis will be in assuming appropriate 
boundary conditions near horizon of black hole. In this approach it 
will turn out to be useful to use the covariant phase space 
formulation of gravity \cite{Crnkovic:1987tz,Julia:2002df}. For this 
reason we shall here review it shortly for any diffeomophism invariant
theory with the Lagrangian $D$-form
\begin{equation}\label{e8}
\mathbf{L}[\Phi]= \mathbfepsilon{}L(\Phi) 
\thickspace\textrm{.}
\end{equation}
Here $\Phi$ denotes collection of fields, $\mathbfepsilon{}$ is the volume $D$-form. Then one can calculate the variation
\begin{equation}\label{e9}
\delta\mathbf{L}[\phi] =\delta \mathbf{E}[\phi] \delta \phi + d\mathbf{\Theta}[\phi, \delta \phi]
\thickspace\textrm{.}
\end{equation}
The $(D-1)$-form, called symplectic potential for Lagrangians of type \refb{e6} was shown in \cite{Iyer:1994ys} to be 
\begin{equation}\label{e10}
\Theta_{pa_1,...a_{n-2}}=2\epsilon_{apa_1...a_{n-2}}(E^{abcd}\nabla _d\delta g_{bc}-\nabla _d E^{abcd} \delta  g_{bc})
\thickspace\textrm{.}
\end{equation}
To any vector field $\xi$ we can associate a Noether current $(D-1)$-form
\begin{equation}\label{e11}
\mathbf{J}[\xi] = 
\mathbf{\Theta}[\phi, \mathcal{L}_\xi \phi] - \xi\cdot\mathbf{L}
\thickspace\textrm{,}
\end{equation}
and the Noether charge $(D-2)$-form
\begin{equation}\label{e12}
\mathbf{J}=d\mathbf{Q}
\thickspace\textrm{.}
\end{equation}
For all diffeomorphism invariant theories the Hamiltonian is a pure surface term \cite{Iyer:1994ys}
\begin{equation}\label{e13}
\delta H[\xi] = \int_{\partial C}(\delta\mathbf{Q}[\xi] - \xi\cdot\mathbf{\Theta}[\phi, \delta \phi])
\thickspace\textrm{.}
\end{equation}
The integrability condition requires that a $(D-1)$-form $\mathbf{B}$ exists with the property 

\begin{equation}\label{e14}
\delta \int_{\partial C}\xi \cdot\mathbf{B}=\int_{\partial C} \xi\cdot\mathbf{\Theta}
\thickspace\textrm{,}
\end{equation}
where $C$ is a Cauchy surface. Then \refb{e13} can be integrated to give
\begin{equation}\label{e14prime}
H[\xi]= \int_{\partial C} (\mathbf{Q}[\xi] - \xi\cdot\mathbf{B})
\thickspace\textrm{.}
\end{equation}
As bulk terms of $H$ vanish, variation of $H[\xi]$ is equal to variations of boundary term
$J[\xi]$. As explained in \cite{Carlip:1999cy,Brown:nw},
 that enables one to obtain the Dirac bracket $\lbrace J[\xi_{1}], J[\xi_{2}] \rbrace_D$

\begin{equation}\label{e15}
\lbrace J[\xi_{1}], J[\xi_{2}] \rbrace_D=
\int_{\partial C}(\xi_{2} \cdot
\mathbf{\Theta}[\phi, \mathcal{L}_{\xi_{1}} \phi]-\xi_{1} \cdot\mathbf{\Theta}[\phi, \mathcal{L}_{\xi_{2}} \phi]-\xi_2 \cdot (\xi_1 \cdot \mathbf{L}))
\thickspace\textrm{,}
\end{equation}
and the algebra

\begin{equation}\label{e16}
\lbrace J[\xi_{1}], J[\xi_{2}] \rbrace_D=J[\lbrace\xi_1, \xi_2 \rbrace]+K[\xi_1, \xi_2]
\thickspace\textrm{,}
\end{equation}
with $K$ as central extension.

Using \refb{e10}, we get a more explicit  form


\begin{eqnarray}
\lbrace J[\xi_{1}], J[\xi_{2}] \rbrace_D &=& \int_{\partial C} 
\epsilon_{apa_{1}\cdots a_{n-2}} \left(\, \xi_{2}^{p} 
E^{abcd}\nabla _{d} \delta_{1}g_{bc}-\xi_{1}^{p}\nabla_{d} 
E^{abcd}\delta_{2}g_{bc}-(1\leftrightarrow 2) \right) 
\nonumber \\ 
&& -\, \xi_{2}\cdot 
(\xi_{1}\cdot \mathbf{L}) \thickspace\textrm{.}
\label{e17}
\end{eqnarray}

\section{Boundary conditions on horizon}

The main idea of the second approach mentioned in the introduction is
to impose existence of Killing horizon and a class of boundary 
conditions on it  proposed by Carlip \cite{Carlip:1999cy} for Einstein
gravity. (For alternative discussions of this method see also 
\cite{Kang:2004js,Dreyer:2001py,Park:1999tj,Park:2001zn,Koga:2001vq}).
We shall assume the validity of these boundary conditions also for the
interactions treated in this paper. The Killing horizon in 
$D$-dimensional spacetime $M$ with boundary $\partial M$ has a Killing
vector $\chi^a $ with the property
\begin{equation}\label{e17prime}
\chi^2 = g_{ab}\chi^a\chi^b = 0 \quad \textrm{at} \quad \partial M
\thickspace\textrm{.}
\end{equation}
One defines near horizon spatial vector $\varrho_a$
\begin{equation}\label{e18}
\nabla _{a} \chi ^2 = -2\kappa \varrho_a
\thickspace\textrm{.}
\end{equation}
We require that variations satisfy

\begin{equation}\label{e19}
\frac{\chi^a\chi^b}{\chi^2}\delta g_{ab}\rightarrow0,\quad\chi^at^a\delta g_{ab}\rightarrow0\quad \textrm{as} \quad \chi^2\rightarrow0 
\thickspace\textrm{.}
\end{equation}
Here $\chi^a$ and $\varrho_a$ are kept fixed, $t^a$  is any unit spacelike vector tangent to $\partial M$.
One considers diffeomorphism generated by vector fields
\begin{equation}\label{e20}
\xi^a = T\chi^a + R\varrho^a
\thickspace\textrm{,}
\end{equation}
Boundary conditions together with the closure of algebra imply
\begin{equation}\label{e21}
R=\frac{\chi^2}{\kappa \varrho^2}\chi^a\nabla _a T\quad \textrm{,} \quad \varrho^a\nabla _aT =0
\thickspace\textrm{.}
\end{equation}
An additional requirement will be necessary as already explained in \cite{Carlip:1999cy}. With the help 
of acceleration of an orbit 
           
\begin{equation}\label{e22.1}
a^{a}=\chi^b\nabla _b\chi^a
\thickspace\textrm{,}
\end{equation}
we define
\begin{equation}\label{e22.2}
\hat{\kappa} ^2=-\frac{a^2}{\chi^2}
\thickspace\textrm{.}
\end{equation}
We ask that
\begin{equation}\label{e22.3}
\delta \int_{\partial C}\hat{\epsilon}(\hat{\kappa}-\frac{\varrho}{|\chi|}\kappa)=0
\thickspace\textrm{.}
\end{equation}

This condition will (see last section) guarantee existence of generators $H[\xi]$ and for diffeomorphisms \refb{e20} will imply

\begin{equation}\label{e22.4}
\int_{\partial C}\hat{\epsilon} \dddot{T}=0
\thickspace\textrm{,}
\end{equation}
and for one parameter group of diffeomorphisms the orthogonality relations

\begin{equation}\label{e22.5}
\int_{\partial C} \hat{\epsilon} T_{n}T_{m}   =\delta_{n+m,0}
\thickspace\textrm{.}
\end{equation}
In order to calculate central term from \refb{e16} we shall use equation
\refb{e17} where we shall integrate over $(D-2)$-dimensional surface 
$\mathcal{H}\cap C$ which is the intersection of Killing horizon with 
the Cauchy surface $C$. In addition to Killing vector  $\chi^a$ we 
introduce other future directed null normal
\begin{equation}\label{e23}
N^a = k^a-\alpha\chi^a - t^a                  
\thickspace\textrm{,}
\end{equation}
where $t^a$ is tangent to $\mathcal{H}\cap C$, and

\begin{equation}\label{e24}
k^a = - \frac{(\chi^a - \varrho^a\frac{|\chi|}{|\varrho|})}{\chi^2}
\thickspace\textrm{.}
\end{equation}
In this way
\begin{equation}\label{e25}
\epsilon_{bca_{1}\ldots a_{n-2}}=\epsilon_{a_{1}\cdots a_{n-2}} \eta _{bc}
\thickspace\textrm{,}
\end{equation}
and
\begin{equation}\label{e26}
\eta_{ab}=2\chi_{[b}N_{c]}=\frac{2}{|\chi|\|\varrho|} \varrho_{[a}\chi_{b]}+t_{[a}\chi_{b]}
\thickspace\textrm{.}
\end{equation}
We proceed now to evaluate the first term of the integrand of \refb{e17} in the leading order in $\chi^2$. Using boundary conditions we can derive the following relation
\begin{equation}\label{e27}
\nabla _{d}\delta g_{ab} = \nabla _{d} \nabla _{a} \xi_{b} + \nabla _{d} \nabla _{b} \xi _{a}=-2\chi_{d}\chi_{a}\chi_{b}\frac{\ddot{T}}{\chi_{4}} + 2\chi_{d} \chi_{(a} \varrho_{b)}(\frac{\dddot{T}}{\kappa \chi^{2}\varrho^{2}} +2 \frac{\kappa \ddot{T}}{\chi^4})
\thickspace\textrm{.}
\end{equation}
After a straightforward calculation and using symmetries  of  $E^{abcd}$  which are those of Riemann tensor we obtain for the first term in \refb{e17}
\begin{equation}\label{e28}
\frac{1}{2} E^{abcd}\eta_{ab}\eta_{cd} (2\kappa T_{2} \dot{T}_{1}-\frac{T_{2}\dddot{T}_{1}}{\kappa})
-(1\leftrightarrow 2) +O(\chi^2)
\thickspace\textrm{.}
\end{equation}
In fact this is the main contribution because we shall show that other terms near horizon are of the order of   
$ \chi^2 $. That is obvious for the third term because Lagrangian is expected to be finite on horizon. The second term after using \refb{e20} and \refb{e26} reads 

\begin{equation}\label{e29}
\frac{\chi_{[a}\varrho_{b]}}{\kappa\varrho^2}[\frac{1}{\kappa}
(\dot{T}_{2}\ddot{T}_{1}-\dot{T}_{1}\ddot{T
}_{2})\chi_{c} -(T_{1}\ddot{T}_{2}-T_{2}\ddot{T}_{1})\varrho_{c}]\nabla _{d} E^{abcd}
\thickspace\textrm{.}
\end{equation}
We want to exploit the fact that $\chi$ is a Killing vector. For this purpose it would be desirable to connect $\nabla_{d}$ with
$\nabla_{\chi}$. We assume that ``spatial''  derivatives are $O(\chi^2)$ near horizon (see Appendix~A of \cite{Carlip:1999cy}), which implies
 \begin{equation}\label{e30}
\nabla _{d} E^{abcd} = (\frac{\chi _{d}\nabla _{\chi}}{\chi^2}+\frac{\varrho_{d}\nabla _{\varrho}}{\varrho^2})
E^{abcd} + O(\chi^2)
\thickspace\textrm{.}
\end{equation}
From \refb{e24}
\begin{equation}\label{e31}
\nabla _{\varrho}= \frac{|\varrho|}{|\chi|}\nabla _{\chi} - |\varrho||\chi|\nabla _{k}
\thickspace\textrm{.}
\end{equation}
This last equation because of consistency with \refb{e21} implies
\begin{equation}\label{e32}
\nabla_{d} E^{abcd} = \frac{\chi_{d} - \varrho_{d}}{\chi^2}\nabla _{\chi}E^{abcd} + O(\chi^2)
\thickspace\textrm{.}
\end{equation}
We are able now to exploit the existence of Killing vector
\begin{equation}\label{e33}
\mathcal{L}_{\chi} E^{abcd} = 0
\thickspace\textrm{,}
\end{equation}
or
\begin{equation}\label{e34}
\nabla _{\chi} E^{abcd} - E^{fbcd}\nabla _{f}\chi^{a} - E^{afcd}\nabla _{f}\chi^{b}
-E^{abfd}\nabla _{f}\chi^{c}-E^{afcf}\nabla _{f}\chi^{d}= 0
\thickspace\textrm{,}
\end{equation}
But due to our boundary conditions  up to leading terms in $\chi^{2}$
\begin{equation}\label{e35}
\nabla ^{a} \chi^{b} = \frac{\kappa}{\chi^2}(\chi^{a}\varrho^{b} -\chi^{b}\varrho^{a}) 
\thickspace\textrm{,}
\end{equation}
we get
\begin{eqnarray}\label{e36}
\nabla _{\chi} E^{abcd} &=&
\frac{\chi_{[a}\varrho_{b]}}{\varrho^6}(\alpha \chi_{c} +\beta \varrho_{c})
(\chi_{d} - \varrho_{d})\times\nonumber\\
& &\times (\chi_{f}\varrho^{a} E^{fbcd}-\chi^{a}\varrho_{f} E^{fbcd}-
\chi_{f}\varrho^{b} E^{facd}
+\chi^{b}\varrho_{f} E^{facd}\nonumber\\
&& ~~ + \chi_{f}\varrho^{c} E^{fdab}-\chi^{c}\varrho_{f} E^{fdab}
+\chi_{f}\varrho^{d} E^{fcba}-\chi^{d}\varrho_{f} E^{fcba})
\;.
\end{eqnarray}
Here
\begin{equation}\label{e37}
\alpha=\frac{1}{\kappa}(\dot{T}_{2}\ddot{T}_{1}-\dot{T}_{1}\ddot{T}_{2}),
\quad
\beta=-(T_{1}\ddot{T_{2}}-T_{2}\ddot{T}_{1})
\thickspace\textrm{,}
\end{equation}
After multiplication we find two classes of terms. One class contains terms  like
\begin{equation}\label{e37.5}
\frac{1}{\chi^2\varrho^2}\chi_{e}\varrho_{f}\chi_{g}\varrho_{h} E^{efgh}=\frac{1}{\chi^2\varrho^2}\chi_{[a}\varrho_{b]}\chi_{[c}\varrho_{d]} E^{abcd}
=\frac{1}{4}\eta_{ab}\eta_{cd} E^{abcd}
\thickspace\textrm{,}
\end{equation}
and such terms are finite but come always in pairs and cancel. All other terms are of the form\\
\begin{displaymath}
\frac{1}{\chi^{4}}\chi_{a}\chi_{b}\varrho_{c}\varrho_{d} E^{abcd}
\thickspace\textrm{,}
\end{displaymath}
and due to antisymmetry properties of  $E^{abcd}$ they vanish. 
We conclude that only first term in \refb{e17} contributes to
$ \lbrace J[\xi_{1}], J[\xi_{2}]\rbrace_{D}$.
Thus after antisymmetrizing in 1 and 2 we obtain
\begin{eqnarray}
\lbrace J[\xi_{1}], J[\xi_{2}]\rbrace_{D} &=& \frac{1}{2} 
\int_{\mathcal{H}\cap C}\hat{\epsilon }_{a_{1}\ldots a_{n-2}}
E^{abcd}\eta_{ab}\eta_{cd} \times
\nonumber \\
&& \qquad\qquad \times \left[ \frac{1}{\kappa}(T_{1}\dddot{T}_{2}
- T_{2}\dddot{T}_1) 
- 2\kappa (T_{1}\dot{T}_{2} -T_{2}\dot{T}_{1})\right]
\thickspace\textrm{.}
\label{e38}
\end{eqnarray}
The Noether charge
\begin{equation}\label{e39}
Q_{c_{3} \ldots c_{n}}=- E^{abcd}\epsilon_{abc_{3} \ldots c_{n}}\nabla _{[c}\xi_{d]}
\thickspace\textrm{,}
\end{equation}
becomes after similar calculation
\begin{equation}\label{e40}
Q_{c_{3} \ldots c_{n}}=-\frac{1}{2} E^{abcd}\eta_{ab}\eta_{cd}(2\kappa T -\frac{\ddot{T}} {\kappa})
\hat{\epsilon} _{c_{3}\ldots c_{n}}
\thickspace\textrm{,}
\end{equation}
which gives us
\begin{eqnarray}\label{e41}
J[\lbrace \xi_{1}, \xi_{2}\rbrace] &=& -\frac{1}{2} 
\int_{\mathcal{H}\cap C} \hat{\epsilon }_{a_{1}\ldots a_{n-2}}
E^{abcd}\eta_{ab}\eta_{cd} \times \\
&& \times [2\kappa({T_{1}}\dot{T}_{2} - {T_{2}}\dot{T}_{1}) 
-\frac{1}{\kappa} 
(\dot{T}_{1}\ddot{T}_{2} - \ddot{T}_{1}\dot{T}_{2} + 
      T_{1}\dddot{T}_{2} - \dddot{T}_{1}T_{2})]
\thickspace\textrm{.}
\end{eqnarray}
From \refb{e38}, \refb{e41} and \refb{e16} follows central charge
\begin{equation}\label{e42}
K[\xi_{1}, \xi_{2}] =-\frac{1}{2} \int_{\mathcal{H}\cap C}
\hat{\epsilon }_{a_{1}\ldots a_{n-2}}E^{abcd}\eta_{ab}\eta_{cd}
\frac{1}{\kappa} (\dot{T}_{1}\ddot{T}_{2}-\ddot{T}_{1}\dot{T}_{2})
\thickspace\textrm{.}
\end{equation}

\section{Entropy and Virasoro algebra}

The main idea is that constraint algebra \refb{e16} can be connected 
to the Virasoro algebra of diffeomorphisns of the real line. For that 
purpose we need to introduce another condition. Denote with $v$ the 
parameter of orbits of the Killing vector
\begin{equation}\label{e43}
\chi^{a}\nabla_{a} v= 1
\thickspace\textrm{.}
\end{equation}
Let us consider functions $T_{1}$, $T_{2}$ of $v$ and ``Killing 
angular coordinates'' $\theta_{i}$ on horizon such that they satisfy
\begin{equation}\label{e44}
\frac{1}{A}\int_{\mathcal{H}\cap C}\hat{\epsilon }T_{1}(v, \theta)
T_{2}(v, \theta)=\frac{\kappa'}{2\pi}\int dv T_{1}(v, \theta)
T_{2}(v, \theta)
\thickspace\textrm{.}
\end{equation}
Here $A=\int_{\mathcal{H}\cap C} \hat{\epsilon }$ is the area of the horizon and $\frac{2\pi}{\kappa'}$ is the period in variable $v$ of functions  $T(v, \theta)$. In particular for rotating black holes
\begin{equation}\label{e45}
\chi^{a} = t^{a} + \sum_{i} \Omega_{i}\psi_{i}^{a}
\thickspace\textrm{,}
\end{equation}
where $t^{a} $ is time translation Killing vector, $\psi_{i}^{a} $ are rotational Killing vectors with corresponding angles $\psi_{i}$ and angular velocities $\Omega_{i}$. We shall sometimes, instead of variables  $t, \psi_{i}$  connected with orbits of $t^{a}, \psi_{i}^{a}$, work with variables $(v, \theta_{i})$ connected with orbits of $\chi_{a}, \theta_{i}^{a}=\psi_{i}^{a}$. Then $v=t$, $\theta_{i}=\psi_{i}-\Omega_{i} v $, 
and we choose for diffeomorphism defining functions $T_{n}$
\begin{equation}\label{e47}
T_{n} = \frac{1}{\kappa'} e^{in(\kappa' v + \sum l_{i}\theta_{i})}
\thickspace\textrm{,}
\end{equation}
where $l_i$ are integers. These functions are of the form

\begin{equation}\label{e48}
T_{n}(v, \theta)=\frac{1}{\kappa'} e^{in\kappa'v}f_{n}(\theta_{i})
\thickspace\textrm{.}
\end{equation}
They satisfy

\begin{equation}\label{e49}
\frac{1}{A}\int \hat{\epsilon}T_{n}T_{m} = \delta_{n+m,0} \frac{1}{\kappa'^2}
\thickspace\textrm{'}
\end{equation}
and in particular
\begin{equation}\label{e50}
\frac{1}{A}\int \hat{\epsilon}f_{n}f_{m} = \delta_{n+m,0}
\thickspace\textrm{.}
\end{equation}
At this point classical Virasoro condition can be checked in the form

\begin{equation}\label{e51}
\lbrace{ T_{m}, T_{n}} \rbrace= -i(m-n)T_{m+n}
\thickspace\textrm{.}
\end{equation}
We also see that condition \refb{e44} is fulfilled and thus enables us to obtain full Virasoro algebra with nontrivial central term $K[T_{m}, T_{n}]$ which can be calculated from \refb{e42}

\begin{equation}\label{e52}
iK[T_{m}, T_{n}]= \frac{\kappa'}{\kappa} \frac{\hat{A}}{8\pi} m^3 \delta_{m+n,0}
\thickspace\textrm{,}
\end{equation}
where
\begin{equation}\label{e52b}
\hat{A} \equiv \frac{1}{8\pi} 
\int_{\mathcal{H}\cap C}\hat{\epsilon }_{a_{1}\ldots a_{n-2}}
E^{abcd}\eta_{ab}\eta_{cd}
\thickspace\textrm{.}
\end{equation}

Here we have used the property that metric does not depend on variables $\theta_{i}$ on which diffeormophism defining functions $T_{n}$ depend. That enabled us to factorize the integral in 
\refb{e42}. Finally, we obtain the Virasoro algebra
\begin{equation}\label{e53}
\lbrace J[\xi_{1}], J[\xi_{2}]\rbrace_{D} = (m-n) J[T_{m+n}] + \frac{c}{12} m^3 \delta_{m+n,0}
\thickspace\textrm{,}
\end{equation}
and central charge is equal to
\begin{equation}\label{e54}
\frac{c}{12} = \frac{\hat{A}}{8\pi}\frac{\kappa'}{\kappa}
\thickspace\textrm{.}
\end{equation}
Now we want to calculate the value of the Hamiltonian. This is given with the first term in relation
\refb{e14prime} where the second term can be neglected\footnote{ As in Einstein case, condition \refb{e22.3}   enables us
 to factorize $\xi\cdot\Theta$ into $E^{abcd}\eta_{ab}\eta_{cd}  \delta(\textrm{terms that vanish on shell)}$, which together with \refb{e14}  implies that $\int \xi\cdot B $ vanishes on shell }.
The first term can be calculated from \refb{e40}
and
\begin{equation}\label{e55}
T_{0} = \frac{1}{\kappa'}
\thickspace\textrm{.}
\end{equation}
Thus
\begin{equation}\label{e56}
J[T_{0}] = -\int_{\mathcal{H}\cap C}
\hat{\epsilon }_{a_{1}\ldots a_{n-2}}E^{abcd}\eta_{ab}\eta_{cd}
\frac{\kappa}{\kappa'}
\thickspace\textrm{,}
\end{equation}
or
\begin{equation}\label{e57}
\Delta  \equiv  J[T_{0}] = \frac{\kappa}{\kappa'} \hat{A}
\thickspace\textrm{.}
\end{equation}
We are now able to use Cardy formula \refb{e5} and obtain following 
expression for entropy
\begin{equation}\label{e58}
S= \frac{\hat{A}}{4}\sqrt{2-\frac{\kappa'}{\kappa}^2}
\thickspace\textrm{.}
\end{equation}
It is remarakable that entropy is proportional to classical classical
entropy with a dimensionless constant of proportionality. We assume
for the period of functions $T_{n}$ the period of the Euclidean black
hole \cite{Carlip:1999cy,Cvitan:2002cs,Barvinsky:2002qu,Kastrup:1996pu,Louko:1996md,Makela:1997rx,Bojowald:1999ex}, which implies
\begin{equation}\label{e59}          
\frac{c}{12} = \Delta
\thickspace\textrm{,}
\end{equation}
and
\begin{equation}\label{e60}    
S= \frac{\hat{A}}{4}= -2\pi\int_{\mathcal{H}\cap C}\hat{\epsilon }_{a_{1}\ldots a_{n-2}}E^{abcd}\eta_{ab}\eta_{cd}
\thickspace\textrm{.}
\end{equation}
As mentioned in the Introduction this derivation is valid for Lagrangian of general form $L=L(g_{ab}, R_{abcd})$.
Let su take the example of Gauss--Bonnet gravity (\ref{igb})
Corresponding tensor $ E^{abcd}$ is then
\begin{equation}\label{e63}   
{E_{ab}}^{cd} = - \Sigma^{[\frac{D}{2}]}_{m=0}m \lambda_{m} \frac{(-)^{m}}{2^{m}}\delta^{cdc_{2}d_{2}\ldots c_{m}d_{m}}_{aba_{2}b_{2}\ldots a_{m}b_{m}}
{R^{a_{2}b_{2}}}_{c_{2}d_{2}}
\ldots 
{R^{a_{m}b_{m}}}_{c_{m}d_{m}}
\thickspace\textrm{.}
\end{equation}
Consequently
\begin{equation}\label{e64}  
S= - 4\pi \Sigma^{[\frac{d}{2}]}_{m}m\lambda_{m}\int \hat{\epsilon} L_{m-1}
\thickspace\textrm{.}
\end{equation}
For the well known case of Einstein gravity
 \begin{equation}\label{e65}  
S=\frac{A}{4}
\thickspace\textrm{,}
\end{equation}
where $A$ is area of black hole horizon.
\section{Conclusion}
We conclude that idea of conformal symmetry near horizon can be useful to interpret the black hole entropy. This idea can be used in two different ways which are both described in this text and they are consistent with  each other when applied to Gauss--Bonnet gravity. The second method, which can be applied in more general cases, consists in  assuming appropriate boundary conditions near Killing horizon. 
One can then identify a subalgebra of diffeomorphism algebra as a 
Virasoro algebra with nontrivial central charge. From Cardy formula 
one can  then determine the entropy. In this way we obtain the 
microscopic interpretation of entropy  i.e.~in terms of the number of
states in Hilbert space. This result can be obtained for special cases
of Einstein gravity \cite{Carlip:1999cy}, Gauss--Bonnet case 
\cite{Cvitan:2002rh,Cvitan:2002cs,Cvitan:2003vq} and for a more 
general class of Lagrangians \cite{Cvitan:2003vq}. It is remarkable
that in all these cases including the general case treated here one 
obtains the classical expression for entropy \cite{Iyer:1994ys}. These
results suggest that conformal symmetry and  Virasoro algebra could 
give further insight in exploring quantum mechanical properties of 
black holes. One is encouraged also to follow this approach due to 
recent proposals for its physical interpretation from the point of 
view of induced gravity \cite{Frolov:2003ed} and an independent 
geometrical interpretation based on properties of the horizon 
\cite{Medved:2004tp,Medved:2004ih}.

\section*{Acknowledgements}

We would like to acknowledge the financial support under the contract
No.\ 0119261 of Ministry of Science and Technology of Republic of 
Croatia.

\end{document}